\documentclass[aps,prl,groupedaddress,superscriptaddress,showpacs,reprint,nofootinbib]{revtex4-2}

\usepackage{amsmath}
\usepackage{amssymb}
\usepackage[final]{changes}
\usepackage{hyperref}
\hypersetup{
    colorlinks=true,
    linkcolor=blue,
    filecolor=blue,      
    urlcolor=blue,
    citecolor=cyan,
}
\usepackage{graphicx}
\usepackage{stackengine}
\usepackage{color}
\usepackage{bm}
\usepackage{physics}
\usepackage{siunitx}
\usepackage[labelformat=simple,labelsep=period,labelfont={sf,bf}]{subfig}
\usepackage{ragged2e}
\usepackage{booktabs}
\usepackage{CJK}
\newcommand{\Jiuzhang}{{\textit{Ji\v{u}zh\={a}ng} }}

\begin{document}


\title{Enhanced Image Recognition Using Gaussian Boson Sampling}


\author{Si-Qiu Gong}
\author{Ming-Cheng Chen}
\author{Hua-Liang Liu}
\author{Hao Su}
\author{Yi-Chao Gu}
\author{Hao-Yang Tang}
\author{Meng-Hao Jia}
\author{Yu-Hao Deng}
\author{Qian Wei}
\author{Hui Wang}
	\affiliation{Hefei National Laboratory for Physical Sciences at Microscale and School of Physical Sciences, 
	New Cornerstone Science Laboratory,
	University of Science and Technology of China, Hefei, Anhui, 230026, China}
	\affiliation{CAS Centre for Excellence and Synergetic Innovation Centre in Quantum Information and Quantum Physics, University of Science and Technology of China, Shanghai, 201315, China}
	\affiliation{Hefei National Laboratory, University of Science and Technology of China, Hefei 230088, China}

\author{Han-Sen Zhong}
\affiliation{Shanghai Artificial Intelligence Laboratory, Shanghai 200030, China}
\affiliation{Shanghai Innovation Institute, Shanghai 200231, China}

\author{Xiao Jiang}
\author{Li Li}
\author{Nai-Le Liu}

\author{Chao-Yang Lu}

    %
\author{Jian-Wei Pan }
    \affiliation{Hefei National Laboratory for Physical Sciences at Microscale and School of Physical Sciences,
	New Cornerstone Science Laboratory,
	University of Science and Technology of China, Hefei, Anhui, 230026, China}
	\affiliation{CAS Centre for Excellence and Synergetic Innovation Centre in Quantum Information and Quantum Physics, University of Science and Technology of China, Shanghai, 201315, China}
	\affiliation{Hefei National Laboratory, University of Science and Technology of China, Hefei 230088, China}




\date{\today}

\begin{abstract}

Gaussian boson sampling (GBS) has emerged as a promising quantum computing paradigm, demonstrating its potential in various applications.
However, most existing works focus on theoretical aspects or simple tasks, with limited exploration of its capabilities in solving real-world practical problems.
In this work, we propose a novel GBS-based image recognition scheme inspired by extreme learning machine (ELM) to enhance the performance of perceptron and implement it using our latest GBS device, \Jiuzhang. 
Our approach utilizes an $8176$-mode temporal-spatial hybrid encoding photonic processor, achieving approximately $2200$ average photon clicks in the quantum computational advantage regime. 
We apply this scheme to classify images from the MNIST and Fashion-MNIST datasets, achieving a testing accuracy of $95.86\%$ on MNIST and $85.95\%$ on Fashion-MNIST. 
These results surpass those of classical method SVC with linear kernel and previous physical ELM-based experiments. 
Additionally, we explore the influence of three hyperparameters and the efficiency of GBS in our experiments. This work not only demonstrates the potential of GBS in real-world machine learning applications but also aims to inspire further advancements in powerful machine learning schemes utilizing GBS technology.
\end{abstract}


\maketitle


Gaussian boson sampling (GBS)\cite{Hamilton2017,Kruse2019} has experimentally demonstrated quantum computational advantage, wherein quantum devices solve specific computational problems overwhelmingly faster than classical computers\cite{Zhong2020,Zhong2021,Madsen2022,Deng2023}.
Quantum computational advantage represents an ongoing competition between classical algorithms and quantum devices.
Recently, we developed a new quantum computer, \Jiuzhang 4.0\cite{liuRobustQuantumComputational}, featuring an $8176$-mode temporal-spatial interferometer to address challenges posed by the latest matrix product state algorithms\cite{Oh2024}.
With significantly more quantum resources than previous iterations, it is intriguing to explore whether this device can tackle practical problems.

There are many potential applications of GBS, including graph-related problems\cite{Arrazola2018,Arrazola2018a,Bradler2018,Deng2023a}, quantum chemistry\cite{Huh2015,Jahangiri2020,Jahangiri2021,shangBosonSamplingEnhanced2024}, and machine learning\cite{Shi2023,sakuraiQuantumOpticalReservoir2025}.
However, due to limitations in the number of modes and programmability, these applications have been restricted to small-scale problems with dozens of modes, such as molecular docking\cite{yuUniversalProgrammableGaussian2023a}.

Machine learning is a promising application of GBS, capable of addressing practical problems involving large-scale data.
Image recognition is a crucial application in machine learning, with uses in autonomous driving\cite{Fujiyoshi2019}, medical diagnosis\cite{Cai2020}, and so on.
It is a classification problem that can be solved by various classical algorithms, including support vector machines (SVM)\cite{Pedregosa2011}, random forests\cite{Pedregosa2011}, and convolutional neural networks (CNN)\cite{Lecun1998}.
These algorithms require the ability to program matrices with more than $10^4$ elements, which is challenging for current GBS devices.
However, some stochastic neural network schemes can reduce the demand for programmability and are suitable for GBS.
Extreme learning machines (ELMs)\cite{Huang2006} and random vector functional-link (RVFL) neural networks\cite{paoLearningGeneralizationCharacteristics1994} are two such stochastic neural networks.
They employ a single hidden layer with random weights to minimize the number of parameters that need to be programmed.
GBS can serve as the hidden layer, with the output layer trained by a classical computer.

In this work, we utilize \Jiuzhang with $1024$ single-mode squeezed states input into an $8176$-mode temporal-spatial hybrid encoding linear optical network, employing ELM and RVFL to classify images from the MNIST\cite{Deng2012} and Fashion-MNIST\cite{xiao2017} datasets.
Here we call them GELM and GRVFL, respectively.
These two models enhance the performance of perceptron and keep to be simple and efficient.
We also test the performance of GBS with coherent states input and show that it will get worse results.
Finally, we benchmark how the classification accuracy is influenced by various hyperparameters, including the number of selected computational bases, principal component analysis (PCA)\cite{Pearson1901,Hotelling1933} features, and samples, as well as the efficiency of the GBS experiment.

The ELM model involves transforming data through a fixed and random nonlinear system, with training occurring at the readout stage.
It nonlinearly transforms input data into a high-dimensional feature space and then employs a linear classifier for classification.
The difference between ELM and RVFL is that the latter merges the original data with the transformed data as the input for the output layer, while ELM uses only the transformed data.
There are some works that implement ELM using physical systems\cite{Suprano2024,ciminiLargescaleQuantumReservoir2025,azamOpticallyAcceleratedExtreme2024,pierangeliPhotonicExtremeLearning2021}.
Some may refer to it as reservoir computing, depending on whether the physical system is designed to process dynamic memory\cite{sakuraiQuantumExtremeReservoir2022a,DeLorenzis2025}.
To train the output layer, we directly use the pseudo-inverse to calculate the weights\cite{Huang2006}.
Therefore, the training of these models is analytical and does not require iterative optimization.
Here, we use GBS to perform the nonlinear transformation of the data.
\begin{figure*}[!htb]
    \centering
    \includegraphics[width=\textwidth]{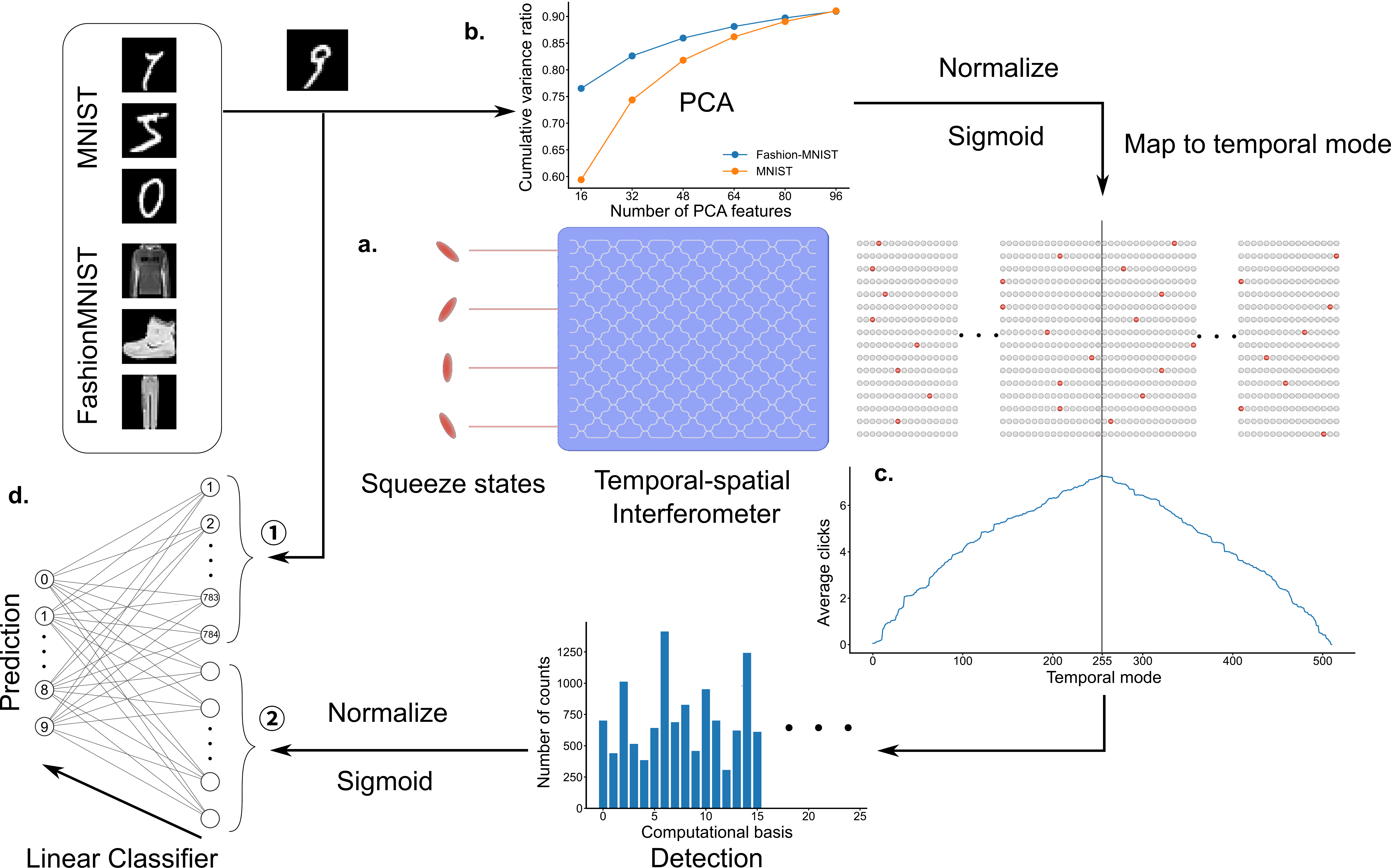}
    \caption{
        \justifying
        The experimental scheme.
        (a) The correspondence between a GBS experiment and selected modes.
        Each PCA feature corresponds to a spatial mode, and the value of the feature determines the temporal mode.
        Columns represent spatial modes, and rows represent temporal modes.
        Selected modes are marked with red circles.
        (b) The cumulative variance ratio of PCA features for MNIST and Fashion-MNIST.
        (c) The trend of the average click number for each temporal mode. 
        It increases in the first $256$ modes and decreases in the last $255$ modes.
        (d) The output layer of the linear classifier.
        The first input part {\normalsize{\textcircled{\scriptsize{1}}}\normalsize} is the original data from MNIST or Fashion-MNIST.
        The second input part {\normalsize{\textcircled{\scriptsize{2}}}\normalsize} is the information of the selected modes from GBS.
        The classifier can be a perceptron, GELM, or GRVFL depending on the input parts.
        If there is only the first input part, it is a perceptron.
        If there is only the second input part, it is a GELM classifier. 
        If both input parts are present, it is a GRVFL classifier.
    }
    \label{fig:scheme}
\end{figure*}
The scheme of our experiment is shown in FIG.~\ref{fig:scheme}.

First, we use \Jiuzhang to generate $9$ million $8176$-mode samples, which consist of $511$ temporal modes and $16$ spatial modes.
These samples will be used as the nonlinear transformation of the input data.
Considering the trend of the average click number for each temporal mode, as shown in FIG.~\ref{fig:scheme}(c), which increases in the first $256$ modes and decreases in the last $255$ modes, we divide the samples into two parts, A and B, starting from the $256$-th temporal mode.

Second, we perform PCA on the MNIST or Fashion-MNIST dataset to reduce their images from $28\times 28$ pixels to $M$ features.
Then, we renormalize the features to a distribution with a mean value of zero and a variance of one, and apply a sigmoid activation function to them.
Finally, we map them to $[0,255]$ or $[0,254]$ depending on the number of temporal modes in the samples.

Third, as shown in FIG.~\ref{fig:scheme}(a), we map the features to GBS mode selection.
Each feature corresponds to a spatial mode, and the value of the feature determines the temporal mode.
For example, if the value of the $i$-th feature is $j$, then the $i$-th spatial mode and $j$-th temporal mode are selected.

Fourth, we count the number of samples on the computational bases for the selected modes.
It is not necessary nor suitable to use all the $2^{16}$ computational bases.
We can select $N$ computational bases with large frequency as follows:
\begin{enumerate}
        \item Randomly select $n_1$ images from the training set.
        \item For each image, count the number of samples on the computational bases for the selected modes.
        \item Select the computational bases with the largest $n_2$ counts.
        \item Select the most frequent $N$ computational bases as the final selected bases.
\end{enumerate} 
Here, we set $n_1 = 1000$ and $n_2 = 2000$.
Considering the structure of our interferometer, we set $M = 16k,  k\in \mathbb{Z^+}$ and partition the features into $k$ groups.
For the fixed selected computational bases, the odd-th groups count samples in partition A, while the even-th groups count samples in partition B.

The merged counts, after renormalization and sigmoid activation function, are used as the second input part {\normalsize{\textcircled{\scriptsize{2}}}\normalsize} of the linear classifier.
Therefore, the second input part is a vector with $kN$ elements.
The first input part {\normalsize{\textcircled{\scriptsize{1}}}\normalsize} is the original data from MNIST or Fashion-MNIST.

Finally, we train the linear classifier using a classical computer and provide predictions for the input images.
The whole model can be a perceptron, GELM, or GRVFL depending on the input parts.
If the classifier only has the first input part, it is a perceptron.
If it only has the second input part, it is a GELM classifier.
If it has both input parts, it is a GRVFL classifier.
\begin{figure*}[!htbp]
    \centering
    \subfloat[Fashion-MNIST,Perceptron,$81.12\%$]
    {
    \includegraphics[width = 0.33\textwidth]{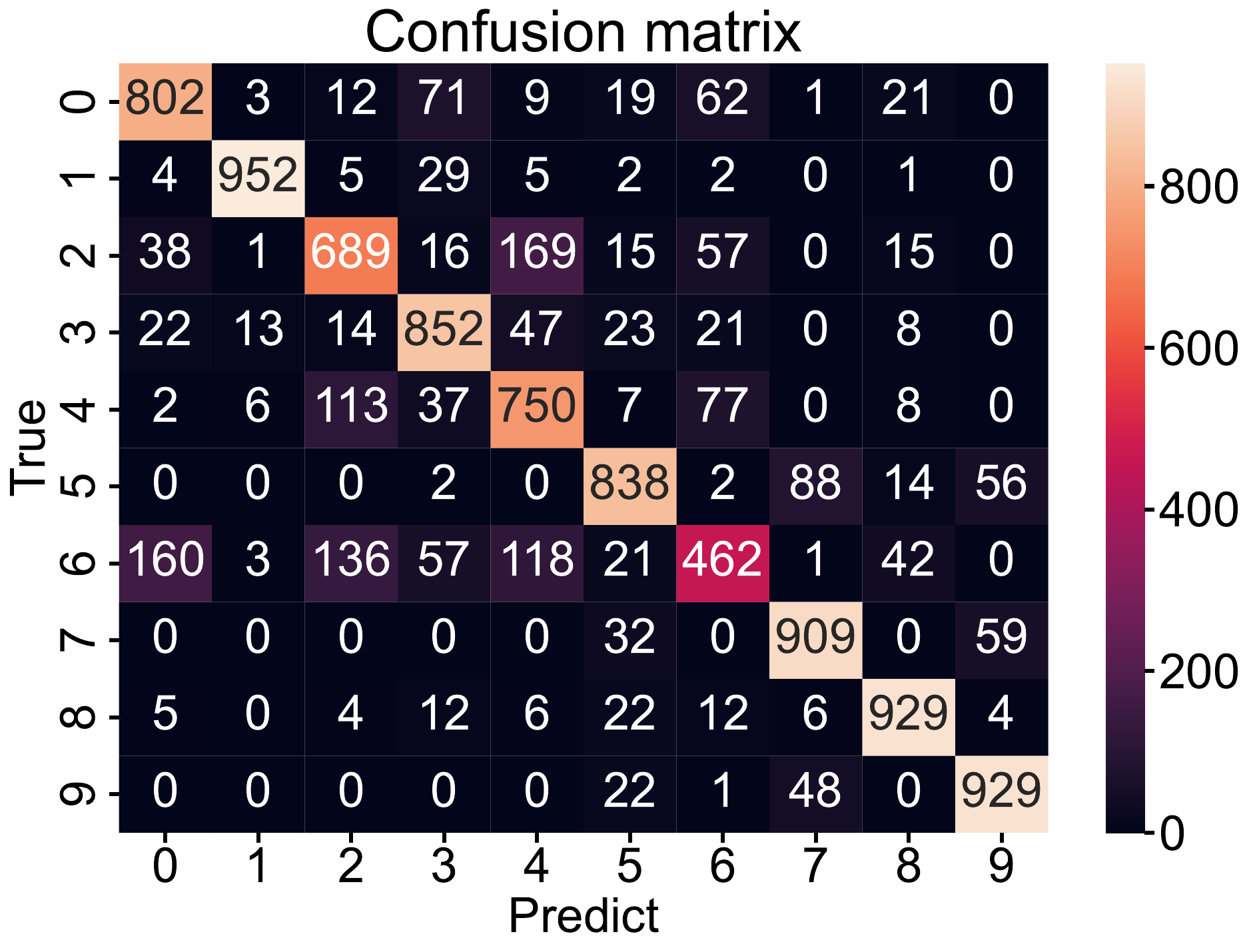}
    }
    \subfloat[Fashion-MNIST,GELM,$85.00\%$]
    {
    \includegraphics[width = 0.33\linewidth]{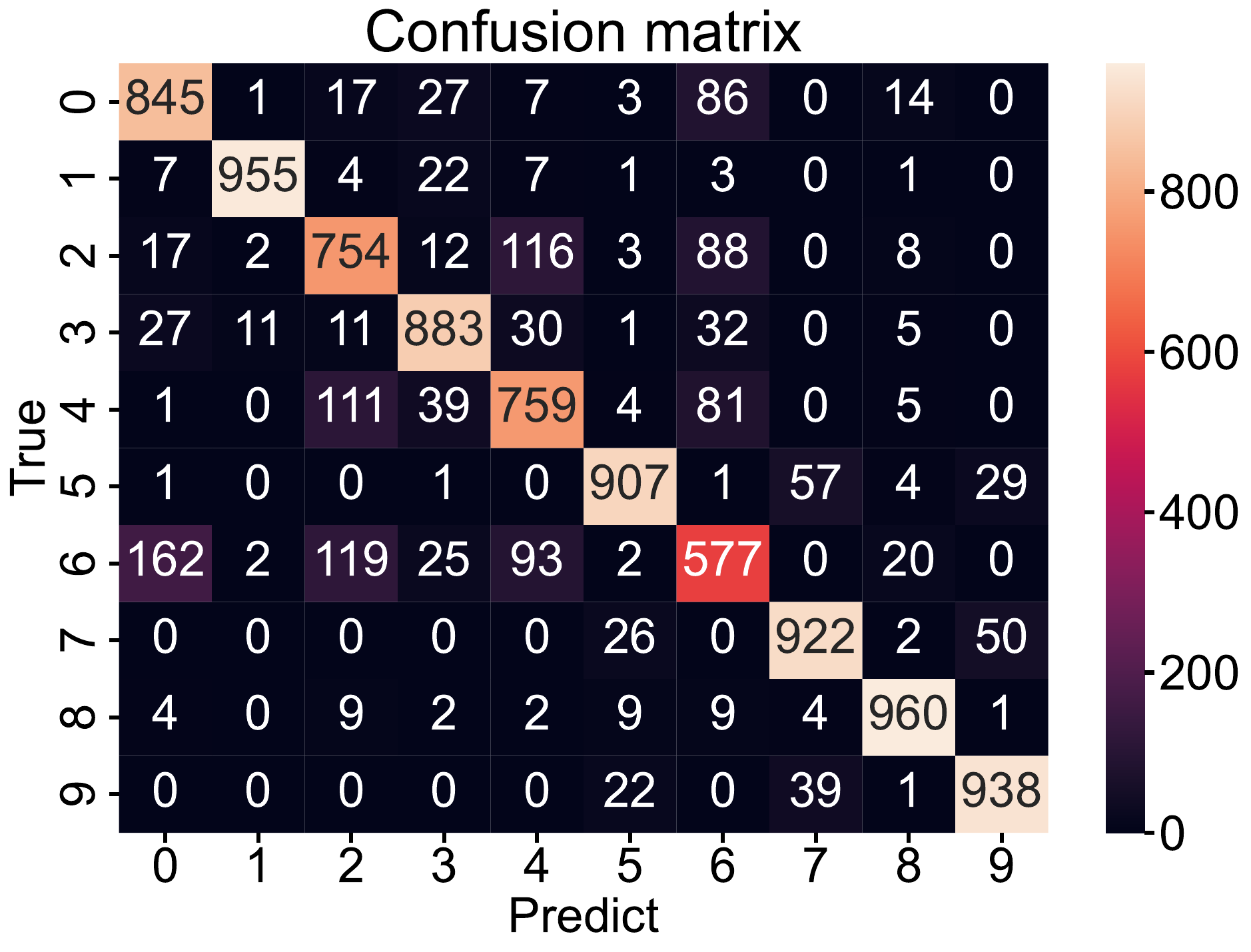}
    }
    \subfloat[Fashion-MNIST,GRVFL,$85.95\%$]
    {
    \includegraphics[width = 0.33\linewidth]{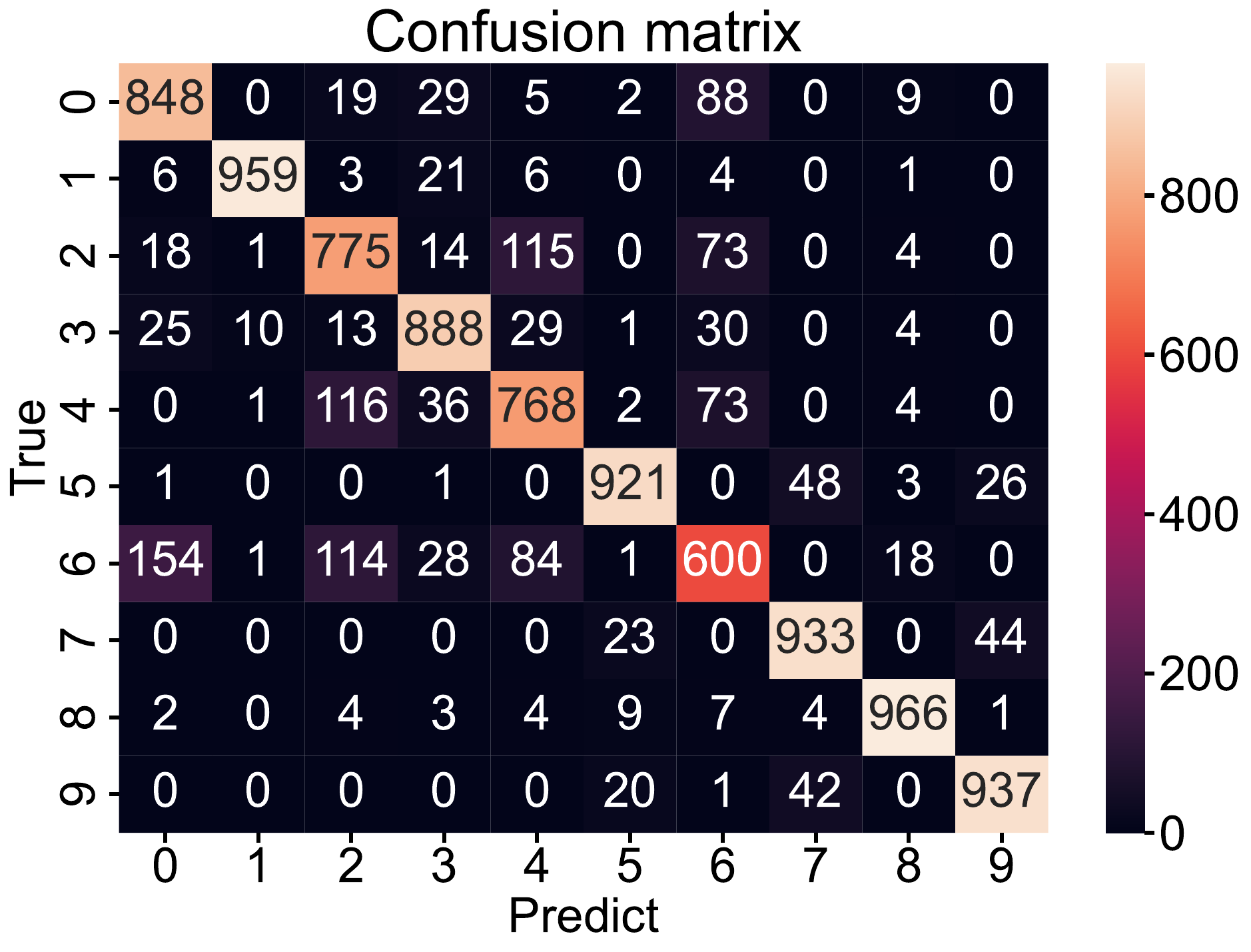}
    }\\
    \subfloat[MNIST,Perceptron,$86.01\%$]
    {
    \includegraphics[width = 0.33\textwidth]{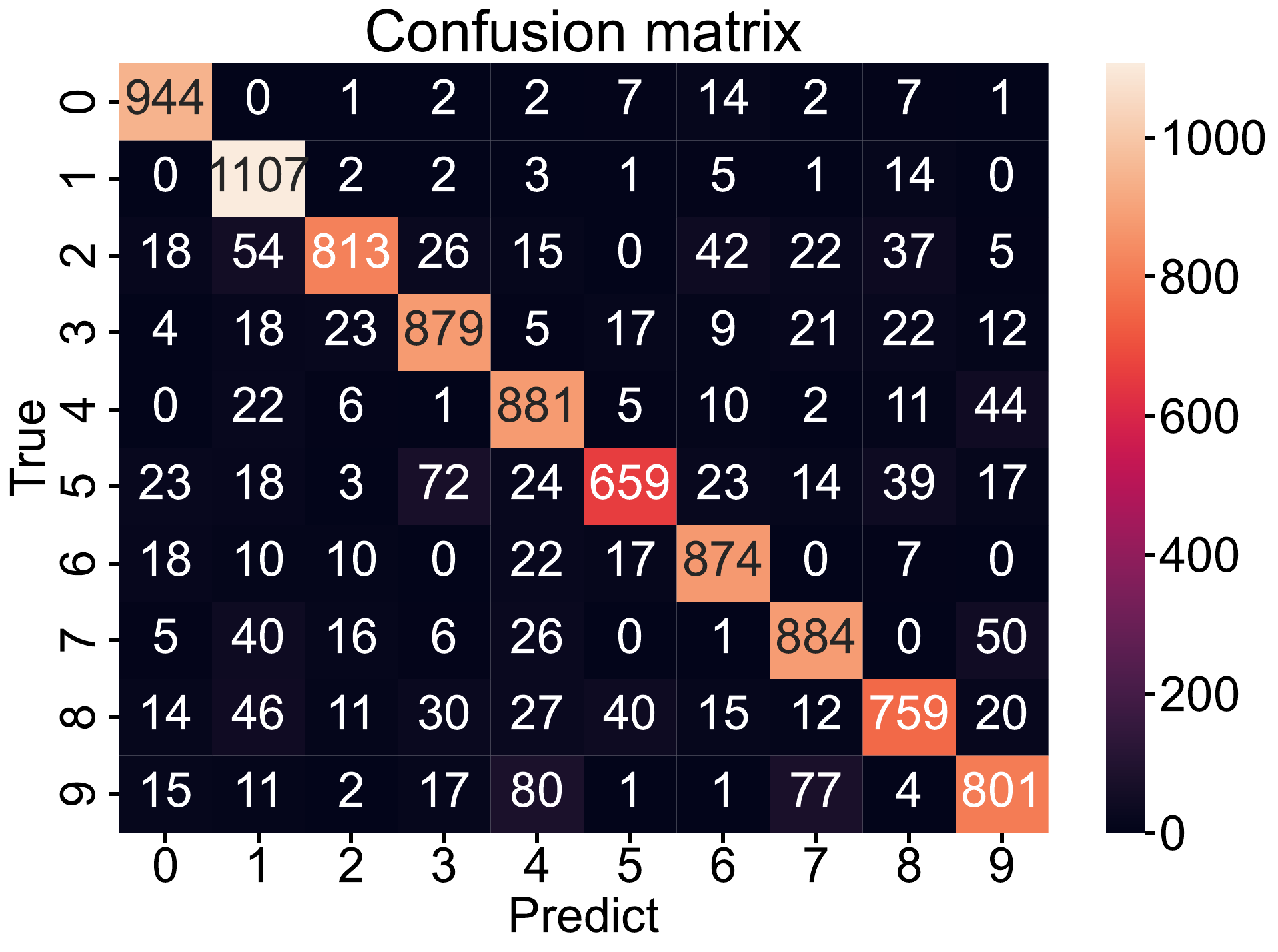}
    }
    \subfloat[MNIST,GELM,$95.27\%$]
    {
    \includegraphics[width = 0.33\linewidth]{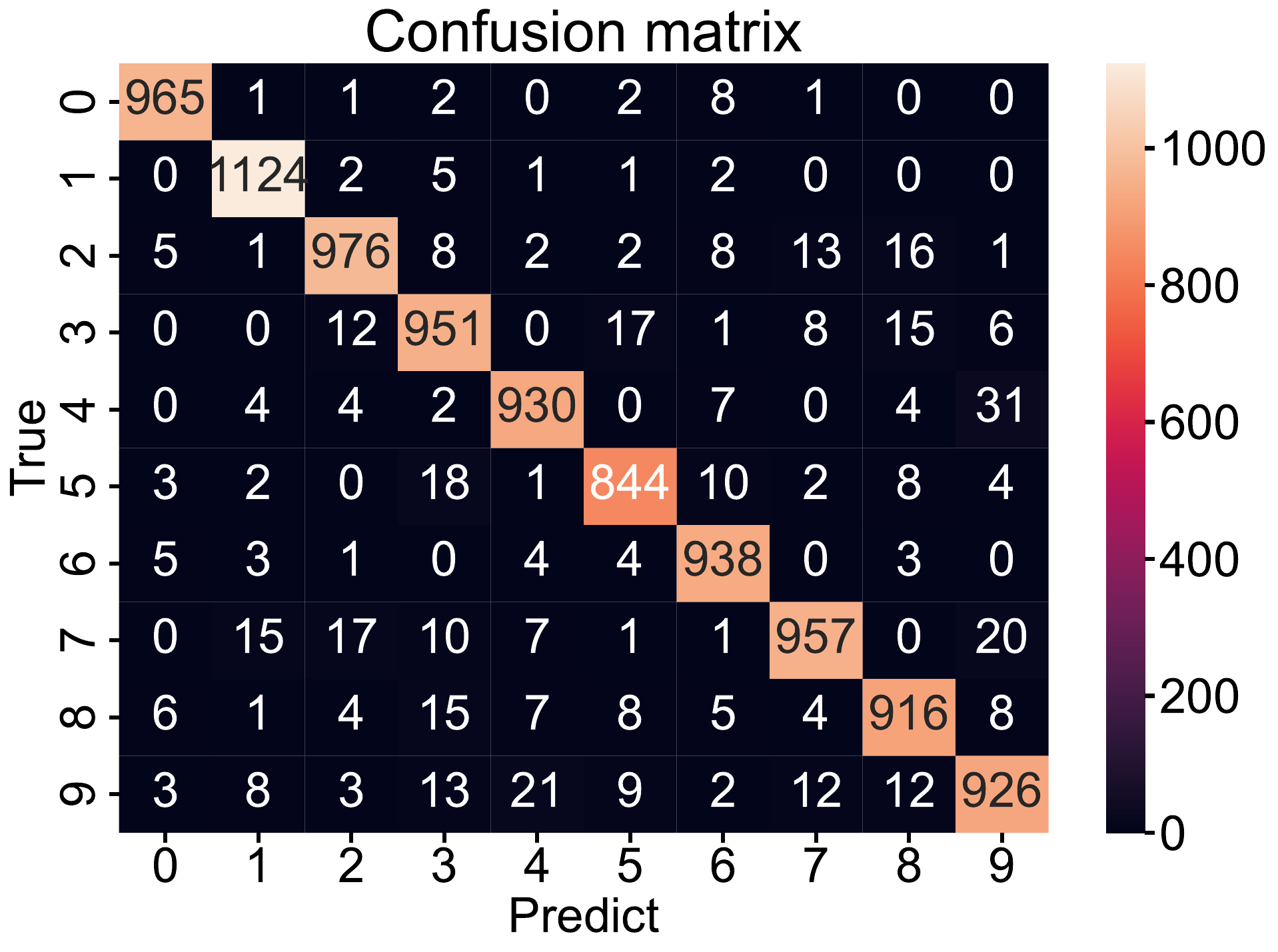}
    }
    \subfloat[MNIST,GRVFL,$95.86\%$]
    {
    \includegraphics[width = 0.33\linewidth]{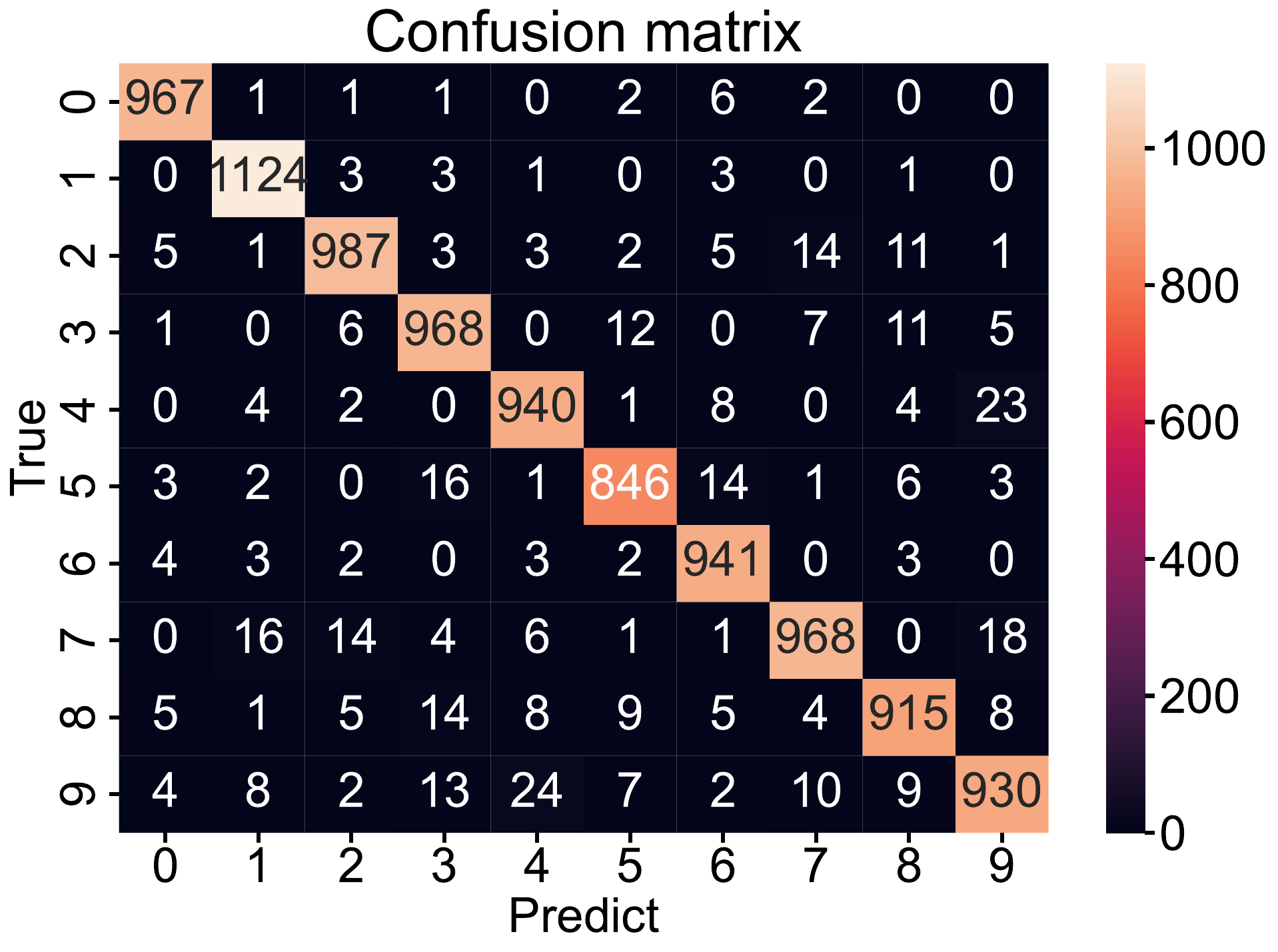}
    }
    \caption{
        \justifying
        Confusion matrices and accuracies for three models and two datasets.
        The number of PCA features is $32$, the number of selected computational bases is $3000$ and the number of GBS samples is $5$ million.
        The first row shows the confusion matrix for the Fashion-MNIST dataset, while the second row shows the confusion matrix for the MNIST dataset.
        The performance of GRVFL is better than GELM and Perceptron.
    }
    \label{fig:Confusion}
\end{figure*}

We use these three types of inputs to train the linear classifier and obtain confusion matrices and testing accuracies for the two datasets, as shown in FIG.~\ref{fig:Confusion}.

The performances of the three models on MNIST are better than those on Fashion-MNIST.
Although the data size and structure of Fashion-MNIST are the same as MNIST, the data in Fashion-MNIST is more complex than that in MNIST.
This complexity makes the classification of Fashion-MNIST more challenging than MNIST.
Therefore, the accuracy of a model on Fashion-MNIST is generally lower than that on MNIST.
Comparing the three models, the test accuracy of GRVFL is the best.
The test accuracy of GRVFL on MNIST is $95.86\%$ and on Fashion-MNIST is $85.95\%$, which are better than $80\%$ results listed in Ref\cite{xiao2017}   and the three physical-based ELM on MNIST\cite{pierangeliPhotonicExtremeLearning2021,azamOpticallyAcceleratedExtreme2024,ciminiLargescaleQuantumReservoir2025}.
\begin{table}[htb]
    \centering
    \begin{tabular}{ccc}
        \toprule
        Model                   &MNIST      &Fashion-MNIST\\
        \midrule
        SVC(C=1 kernel=linear) \cite{xiao2017}              &92.9\%     &83.9\%       \\
        Ref\cite{azamOpticallyAcceleratedExtreme2024}   &93.25\%    &-      \\
        Ref\cite{pierangeliPhotonicExtremeLearning2021}        &92.18\%    &-      \\
        Ref\cite{ciminiLargescaleQuantumReservoir2025}     &$\approx 60\%$     &-      \\
        Coherent-ELM    &75.14\%    &74.84\%      \\
        Coherent-RVFL   &89.98\%    &82.79\%      \\
        Perceptron              &86.01\%    &81.12\% \\
        GELM                 &95.27\%    &85.00\% \\
        GRVFL                &95.86\%    &85.95\% \\
        \bottomrule
    \end{tabular}
    \caption{
        \justifying
        The test accuracy of different models on MNIST and Fashion-MNIST datasets.
    To be fair, we use the same training and testing set as the SVC with linear kernel and three physical-based ELM and do not use $k-$fold cross-validation.
    Because our models are trained analytically and the training and testing sets are unchanged, the results will not change when retraining the model.
    As standard errors were not reported for the three physical-based ELM, we likewise omit them here.
    }
    \label{tab:Accuracy}
\end{table}
The comparison of the test accuracy of different models on MNIST and Fashion-MNIST datasets is shown in Table~\ref{tab:Accuracy}.
More classical results can be found in Ref\cite{xiao2017}.

Our two models that use GBS as the nonlinear transformation of the data outperform the SVC with linear kernel and three physical-based ELM.
To compare the performance of GBS with its classical counterparts, we also generate $5$ million samples with coherent states input  whose $\left\vert \bm{\alpha} \right\vert^2  = \sinh(\bm{r})^2$, i.e., their input average photon number equals to our experiment,  and use them as the nonlinear transformation.
All the other settings remain the same as those of GELM and GRVFL.
These models are called Coherent-ELM and Coherent-RVFL, respectively.
The results are shown in Table~\ref{tab:Accuracy} and are worse than those of GELM and GRVFL.
It indicates that GBS is a better choice than coherent states in transforming data for image classification.

The accuracy of GELM is better than that of the perceptron, indicating that using GBS as the nonlinear transformation of the data to map input data to a high-dimensional feature space is beneficial.

The accuracy of GRVFL is better than that of GELM and the perceptron, suggesting that the original data and the transformed data provide different information for classifying the images.
Combining both types of data is more effective than using only one.
This is also the principle behind broad learning systems\cite{Chen2018}.

These results mean that add GBS to the perceptron can enhance its performance, no matter seriesly or parallelly.

\begin{figure*}[!htbp]
    \centering
    \subfloat[Accuracy vs basis number\newline FashionMNIST]
    {
    \includegraphics[width = 0.24\textwidth]{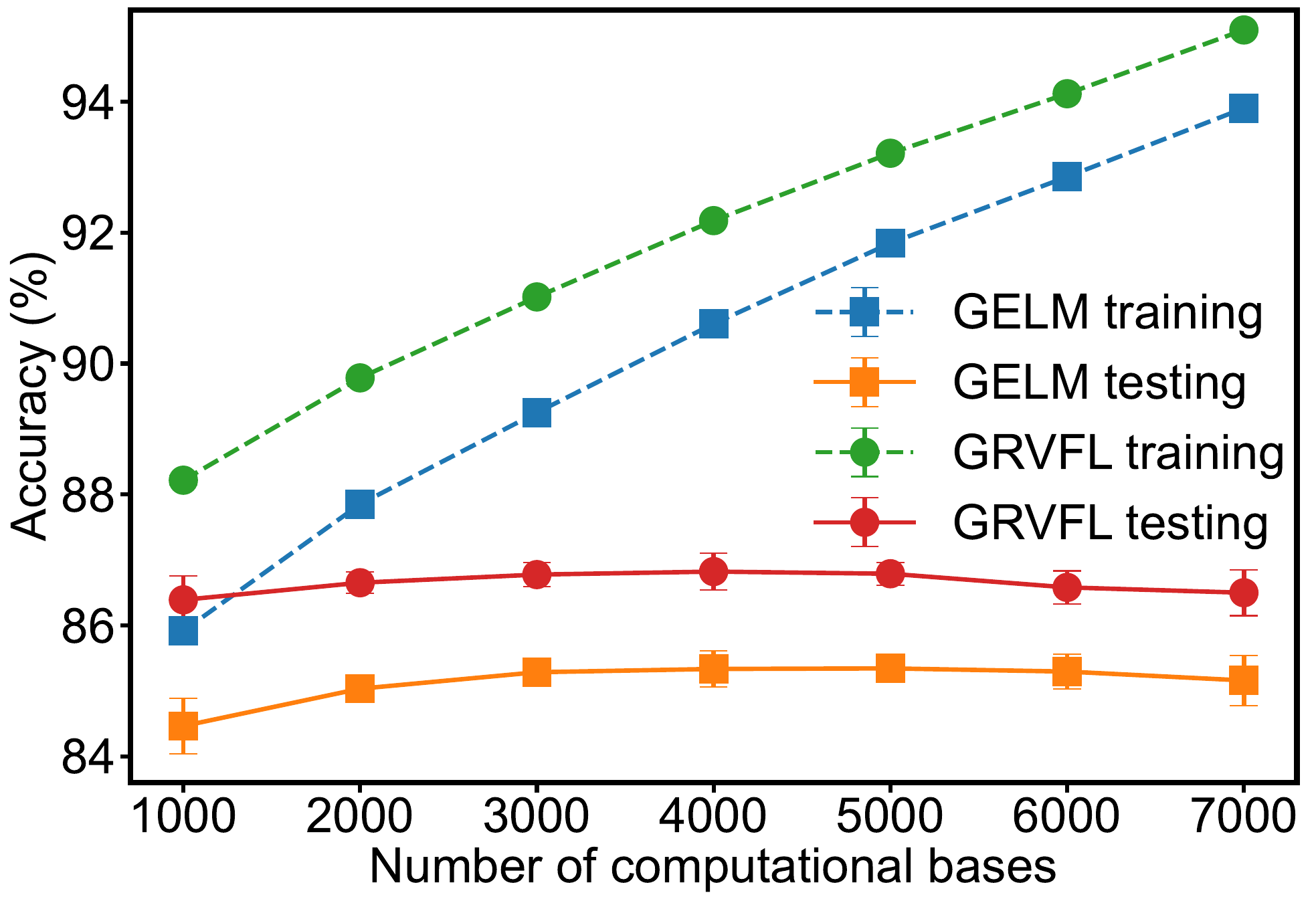}
    }
    \subfloat[Accuracy vs feature number\newline FashionMNIST ]
    {
    \includegraphics[width = 0.25\linewidth]{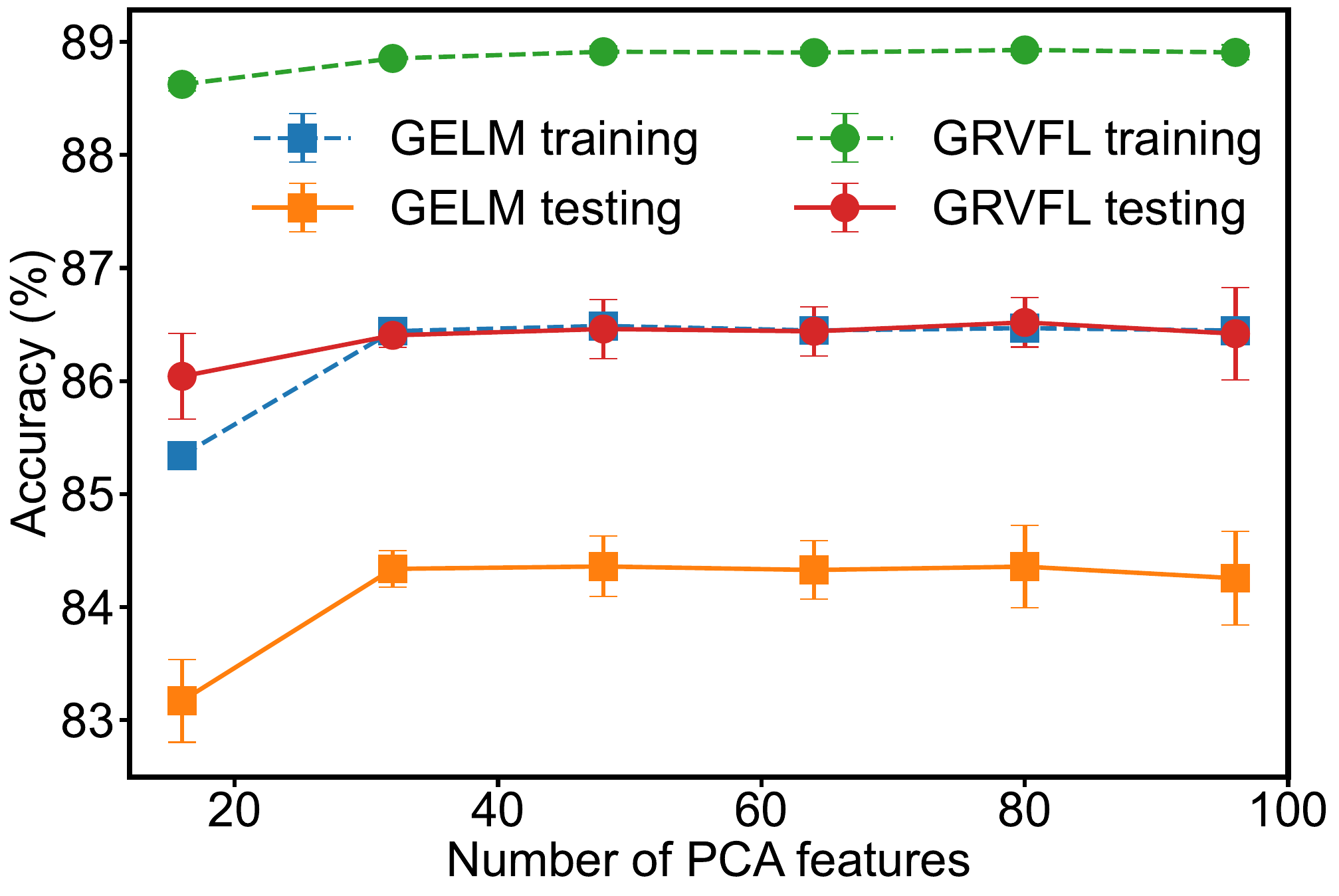}
    }
    \subfloat[Accuracy vs sample number\newline FashionMNIST  ]
    {
    \includegraphics[width = 0.24\linewidth]{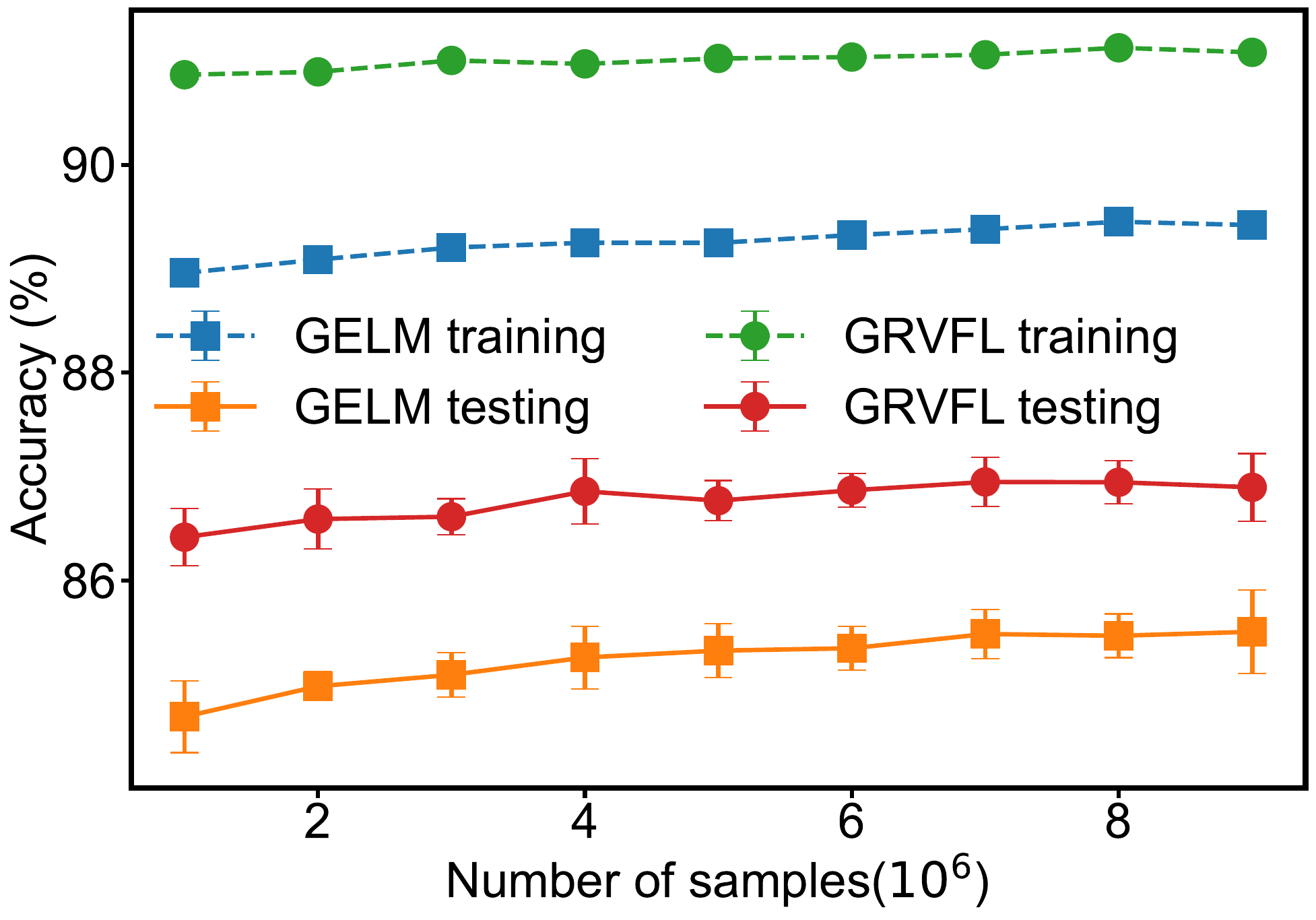}
    }
    \subfloat[Accuracy vs efficiency \newline FashionMNIST  ]
    {
    \includegraphics[width = 0.24\linewidth]{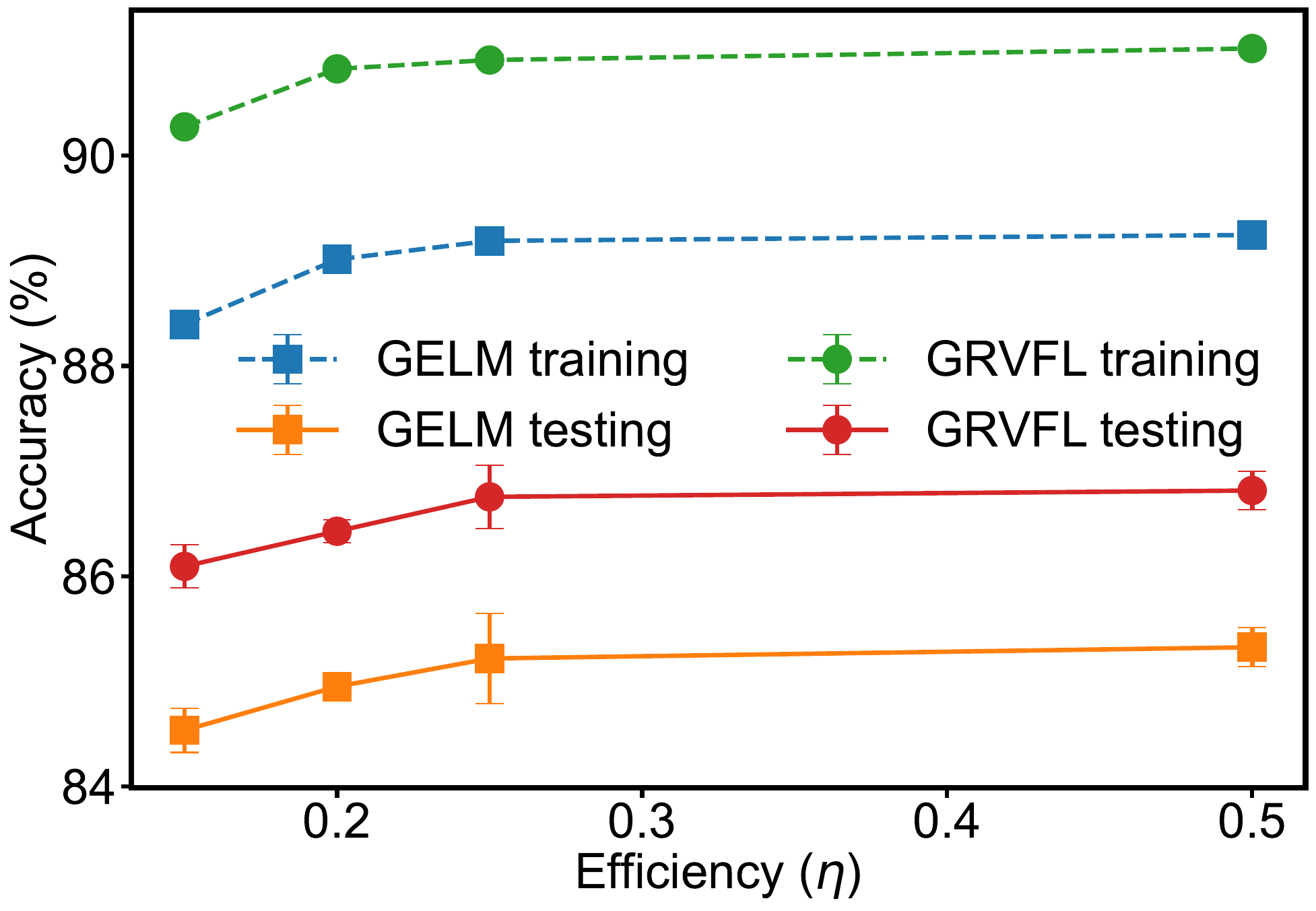}
    }
    \\
    \subfloat[Accuracy vs basis number\newline MNIST ]
    { 
    \includegraphics[width = 0.24\linewidth]{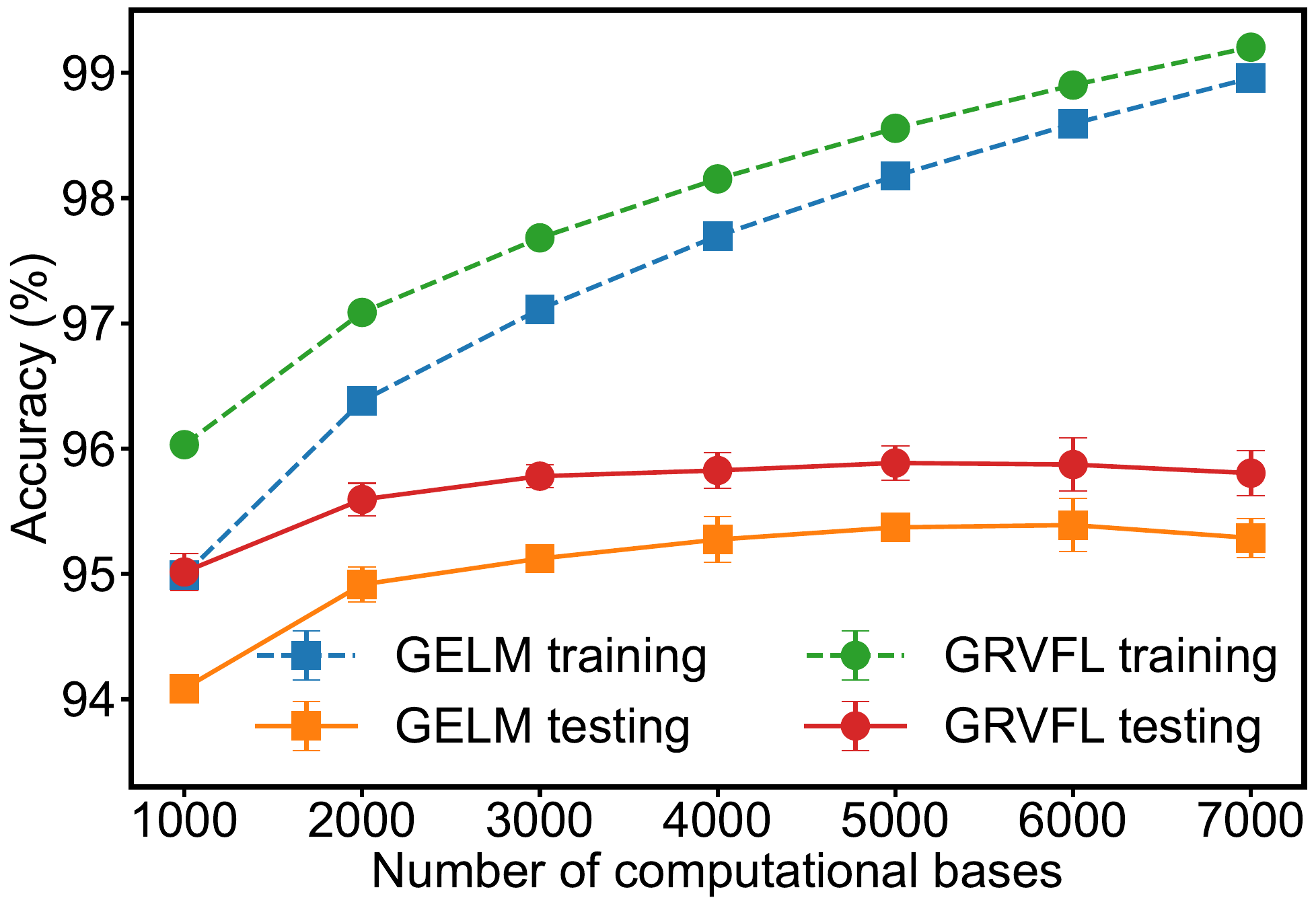}
    }
    \subfloat[Accuracy vs feature number\newline MNIST ]
    {
    \includegraphics[width = 0.25\linewidth]{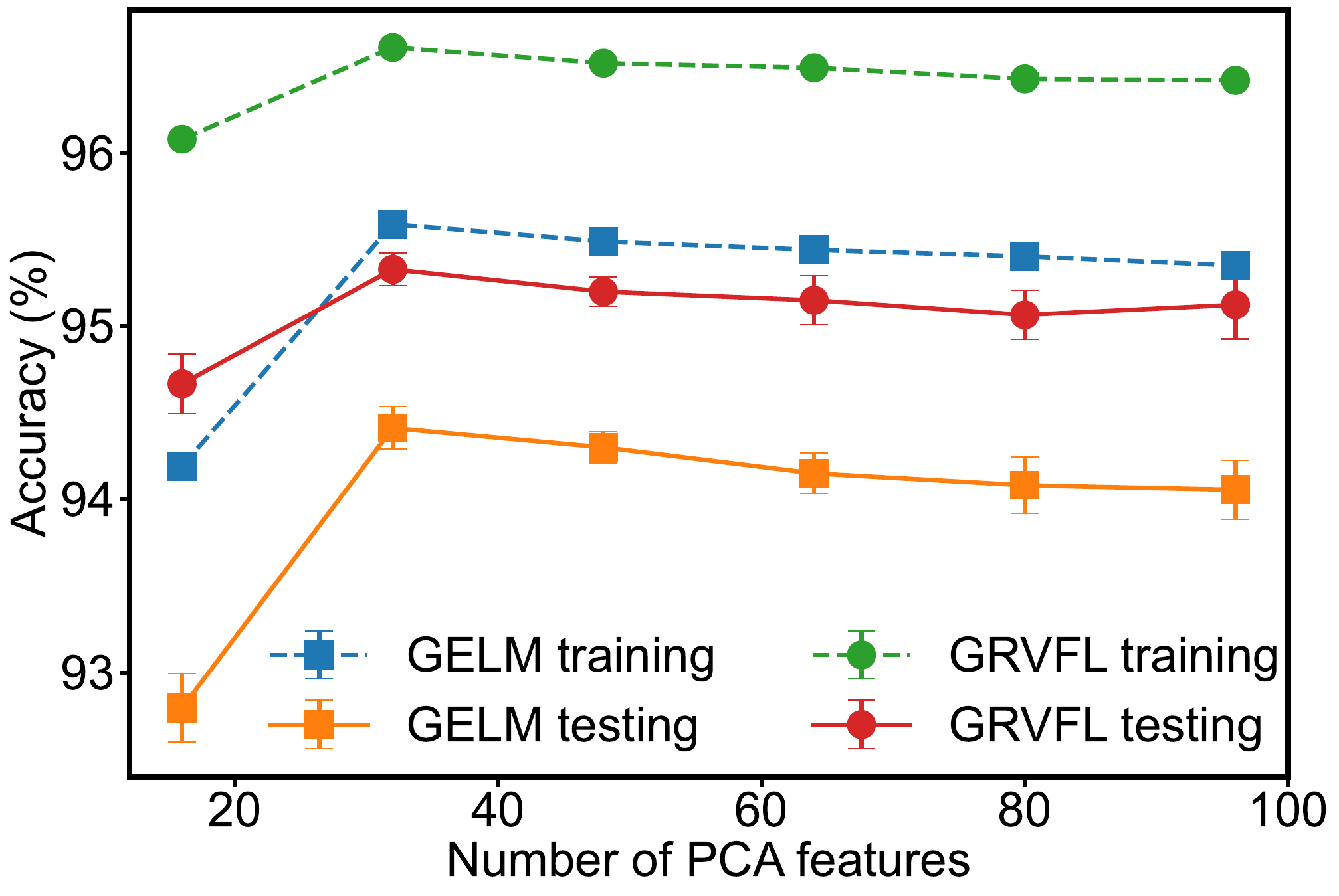}
    }
    \subfloat[Accuracy vs sample number\newline MNIST ]
    {
    \includegraphics[width = 0.24\linewidth]{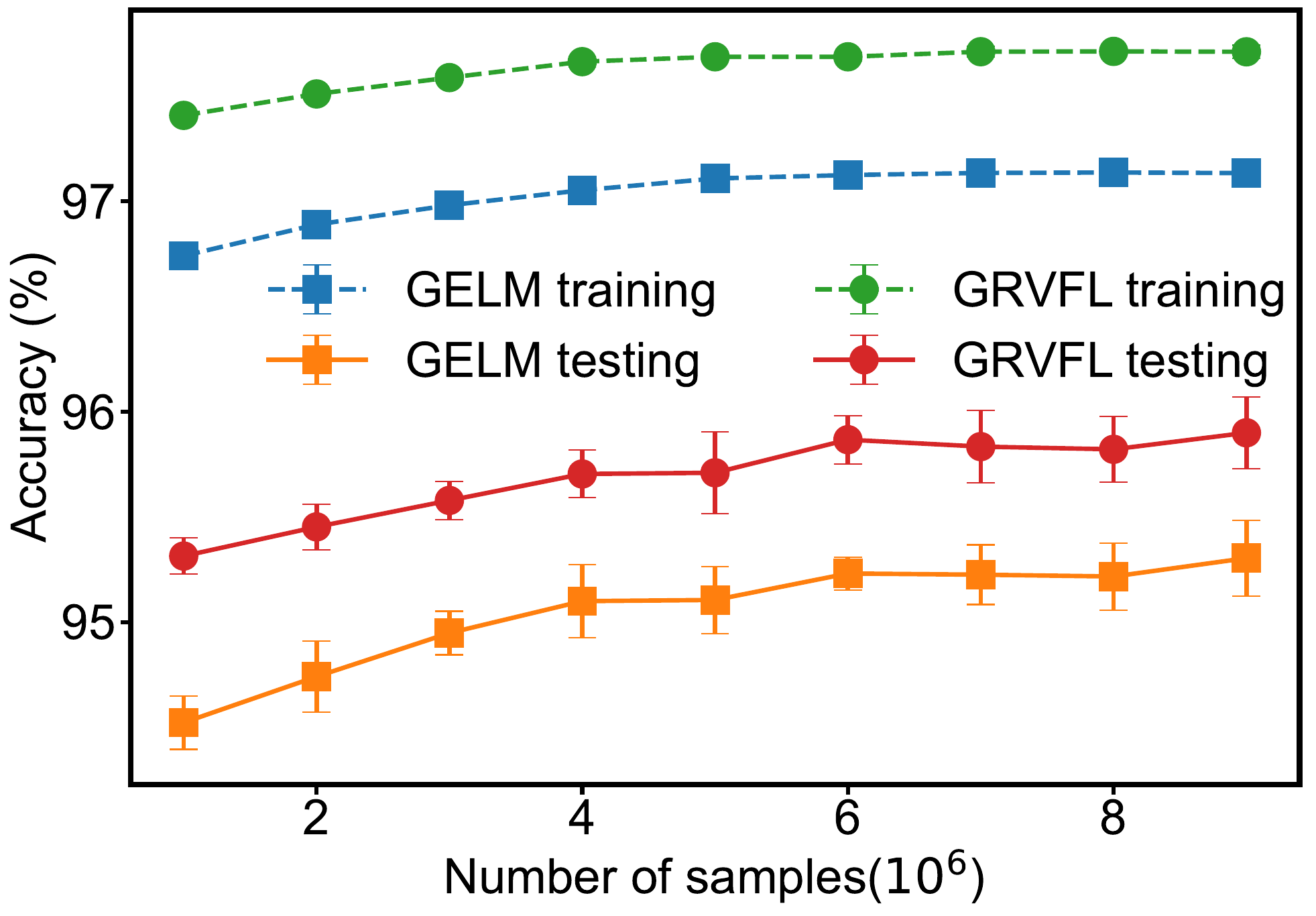}
    }
    \subfloat[Accuracy vs efficiency\newline MNIST ]
    {
    \includegraphics[width = 0.24\linewidth]{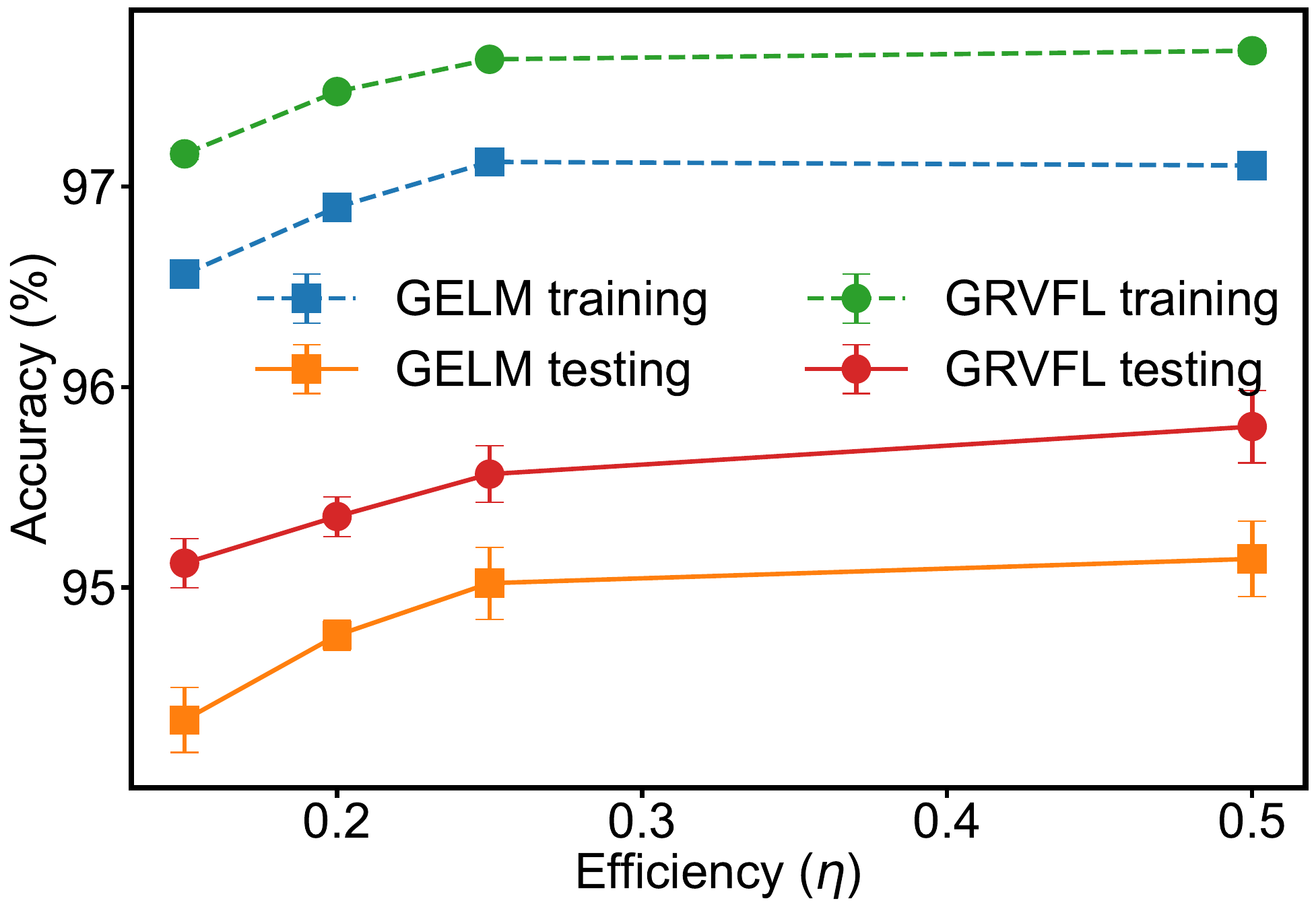}
    }
    \caption{
        \justifying
        The relation between the accuracy and the number of selected computational bases(a,e), the number of PCA features(b,f), the number of GBS samples(c,g) and the system efficiency(d,h) for GELM and GRVFL the two models on FashionMNIST and MNIST.
        The default hyperparameters are the number of selected computational bases $3000$, the number of PCA features $32$ and the number of GBS samples $5$ million.
        And we use the experimental samples with $\eta \approx 0.50$ by default.
        The accuracy is the mean of 7-fold cross validation results.
        Error bar indicates double standard error.
    }
    \label{fig:Performance}
\end{figure*}

There are many hyperparameters in our models, such as the number of selected computational bases, the number of PCA features, and the number of GBS samples.
It is interesting to understand how these hyperparameters influence the performance of the models.
We use $7$-fold cross-validation\cite{kohavi1995study} to evaluate the performance of the models with different hyperparameters.
The results are shown in FIG.~\ref{fig:Performance}, and the error bar indicates double standard error.

The first hyperparameter is the number of selected computational bases.
This hyperparameter determines the dimension of the feature space.
Selecting more computational bases will increase the number of parameters in the model.
Generally, it will increase the training accuracy of the model.
However, it can also lead to overfitting.
As shown in FIG.~\ref{fig:Performance}(a,e), the training accuracies of GELM and GRVFL increase with the number of selected computational bases, while the testing accuracies initially increase and then decrease.
The differences between the training and testing accuracies of GELM and GRVFL also increase, indicating that the model is overfitting when the number of selected computational bases is too large.

The second hyperparameter is the number of PCA features.
PCA can reduce the dimension of the data.
However, it may also result in the loss of some information from the data.

The cumulative variance ratio of PCA features is shown in FIG.~\ref{fig:scheme}(b).
Setting the number of PCA features to $32$ yields a cumulative variance ratio of $0.744(0.826)$ for MNIST(Fashion-MNIST).
The increasing trend of the cumulative variance ratio becomes slower after this point.
To investigate the influence of the number of PCA features on the performance of the models and avoid the influence of the computational bases number, we perform the training as follows.
We set the number of selected computational bases to $3000$ and the number of GBS samples to $5$ million.
Then, we calculate the frequency of the selected computational bases with $n$ groups of $16$ PCA features successively.
The final frequency, used as the input for the linear classifier, is the weighted sum of the frequencies of the $n$ groups.
The weights are the cumulative variance ratio of the $n$ groups.
The relationship between the accuracy and the number of PCA features is shown in FIG.~\ref{fig:Performance}(b,f).
The accuracies increase significantly when the number of PCA features increases from $16$ to $32$, and then there is no obvious change, i.e., small changes or the error bars cover the changes.
This indicates that we can capture most of the information from the data by using $32$ PCA features.

The third hyperparameter is the number of GBS samples.
We use the frequency of the selected computational bases as the mapping result in the high-dimensional space.
The more samples we use, the more accurate the mapping result will be.
However, using more samples also increases the time required to generate the samples and count the results.
The relationship between the accuracy and the number of GBS samples is shown in FIG.~\ref{fig:Performance}(c,g).
The training and testing accuracies of ELM and RVFL both increase with the number of GBS samples.
When the number of GBS samples exceeds $5$ million, the accuracies remain nearly unchanged.
This means that we can achieve good model performance with several million samples, and $5$ million is a reasonable choice.
It represents a balance between performance and time cost.

In GBS, photon loss is a major noise source and is inevitable.
To characterize the influence of photon loss on the performance of our models, we generate $5$ million samples using the same squeezed states input and scale the interferometer with different factors to simulate different efficiencies $\eta$.
The samples with $\eta \approx 0.50$ are experimental samples, while the others are generated using matrix product states (MPS)\cite{Oh2024} with a bond dimension $\chi=1000$.
The results are shown in FIG.~\ref{fig:Performance}(d,h).
We set the number of selected computational bases to $3000$ and the number of PCA features to $32$.
The selected computational bases are fixed for different efficiencies to avoid the influence of computational bases selection.
The results show that the training and testing accuracies of GELM and GRVFL both increase with efficiency.
The testing accuracies of the lowest efficiency samples are still better than those of SVC with linear kernel and the three physical-based ELM.

In this Letter, we use \Jiuzhang to classify images from the MNIST and Fashion-MNIST datasets, enhancing the performance of perceptron and achieving better testing accuracy than SVC with linear kernel and three physical-based ELM.
This work represents a trial to use GBS with $8176$ modes and approximately $2200$ average photon clicks to solve practical problems, outperforming the results using coherent states input and demonstrating its competence in this regard.
Compared to the scheme using an interferometer with dozens of modes\cite{sakuraiQuantumOpticalReservoir2025}, our scheme fully utilizes our mode resources and significantly reduces programmability requirements.
This also reduces the time cost of the experiment.
Our scheme can be integrated into more complex models to address additional practical problems.
We hope that our work can inspire more real-world applications of GBS and leverage quantum properties for practical machine learning tasks.

\bibliography{image_recognition}

\end{document}